\documentstyle[aaspptwo,flushrt]{article}

\input epsf.sty

\def\figinsert#1#2{\epsfbox{#1}\message{#2} }

\begin{document}

\title{Dwarf Galaxies with Gentle Star Formation and the Counts of
Galaxies from the Hubble Deep Field}

\author{ Ana Campos}

\affil{Instituto de Matem\'aticas y F\'{\i}sica Fundamental, CSIC, Spain,  \\
and \\
Observatorio Astron\'omico Nacional, Campus Universitario de Alcal\'a de Henares, \\
Apdo: 1143, E-28800 Alcal\'a de Henares, Madrid, Spain. \\ 
e-mail: ana@oan.es}

\begin{abstract}

In this paper the counts and colors of the faint galaxies observed in
the Hubble Deep Field are fitted by means of simple luminosity
evolution models which incorporate a numerous population of \it fading
\rm dwarfs. The observed color distribution of the very faint galaxies
allows now to put constrains on the star formation history in
dwarfs. It is shown that the star-forming activity in these small
systems has to proceed in a \it gentle \rm way, i.e. through episodes
each one lasting much longer than a simple instantaneous burst of star
formation. By allowing dwarfs to form stars in this \it gentle \rm way
the number of predicted red remnants is severely reduced, in good
agreement with the observations. Then, if the faint counts are to be
fitted by means of dwarfs, the simple model for dwarfs forming stars in
single, very short episodes is challenged, and a more complex star
formation history has to be invoked. Recent observational evidence
supporting this new dwarf models are also discussed.

\end{abstract}


\vspace{2cm}

\begin{center}

To appear in {\em The Astrophysical Journal}, May 1997

\end{center}

\section{Introduction}

The Hubble Deep Field (HDF; Williams et al. 1996) has provided the
deepest view so far achieved from the Universe. Metcalfe et al. (1996)
have computed galaxy counts to the faintest limits (U=26, B=29, R=28
I=28), shown to be in good agreement with ground-based data at brighter
levels (B=28.2). The information on galaxy formation and evolution that
can be obtained from the HDF is striking, as it has been shown during
the past months.

The HDF has a very small field of view (5.3 arcmin$^2$), and so the
dominant population of galaxies is built up by very faint objects.  In
fact, the number of galaxies with apparent magnitude I$\sim 28$ exceeds
that of galaxies with I$\sim 23$ by almost a factor of 20.  Which kind
of galaxies constitutes the bulk of objects in the HDF?

The first result that comes out after a first look to the HDF
counts is that, even if there is some flattening in the deepest
magnitude bins, down to the faintest limits the counts continue to
increase with a quite steep slope. This is a difficult result to
interpret in \it any cosmological setting, \rm not just because of the
cosmological turn-down effect of the volume (especially if $q_0=0.5$)
but also because of the Lyman break being red-shifted into the
different photometric bands (approx z=3, 4, 5 and 6 for U, B, R and I
respectively), hiding any galaxy beyond those limits. An observational
confirmation of the presence of the Lyman break is found in the
Keck-spectra of galaxies at $z\sim 3$ (Steidel, Pettini \& Hamilton
1996). In fact, as I will show in the next sections, the level of
counts is not easy to predict with \it pure \rm luminosity evolution
models, no matter the value of $q_0$ used.

The level of counts at the faint limits can be increased by including
merging in the models, as was first suggested by Guiderdoni and
Rocca-Volmerange (1991) and by Broadhurst, Ellis \& Glazebrook (1992)
for an spatially-flat cosmology. The number-density of galaxies could
increase as we look back toward higher redshifts, galaxies being
splitted up into the fragments that eventually will merge to build up
the present-day population of galaxies.

However, the rate of merging has recently been claimed to be \it
moderate, \rm (eg. Barger et al. 1996; Dickinson 1996; Griffiths et
al. 1996). Ellis et al. (1996) argue that the very little scatter in
the U-V color of ellipticals in clusters at $z\sim 0.5$ is consistent
with the previous suggestion by Bower, Lucey \& Ellis (1992; from their
study of galaxies in the Coma cluster) that ellipticals formed at high
redshift and since then have \it passively \rm evolved. Then, if
ellipticals were assembled by merging of smaller fragments this had to
occur at very early times (Kauffmann 1996).

Some authors (eg. Koo \& Kron 1992) have suggested that the bulk of
faint blue galaxies in the deep counts could be intrinsically faint
galaxies (dwarfs) located at low redshift. Even in an Einstein-de
Sitter (EdS) model, the contribution of a (numerous) population of
dwarfs to the counts will continue to increase with an Euclidean slope
becoming the dominant population at the faint levels, as first noticed
by Driver et al. (1994). Babul \& Ferguson (1996) show that by
including a population of ``fading'' dwarfs the level of counts can be
easily increased to the observed levels without the recourse to
number-density evolution (which, in any case, might exist).

In this paper I try to get new insight into the role of dwarfs to
interpret the counts of faint galaxies, in an attempt to constrain the
star formation process in these low mass systems by means of the galaxy
colors in the HDF. Whereas Babul \& Ferguson assume that the star
formation in dwarfs takes place in single, very short ($\sim 10^7$
years) episodes, I will argue that in order to fit the counts to the
faintest levels by means of dwarfs {\it without over-predicting} the
number of red, faint remnants, the star formation should take place in
a more {\it gentle} way. In fact, as discussed below, analysis of the
photometry of individual stars in nearby dwarfs (Smecker-Hane et
al. 1996) show that dwarfs go through episodic bursts each one lasting
$\sim$ 1 Gyr.

The modeling of bright and dwarf galaxies to fit the counts is
presented in \S 2. \S 3 is devoted to the comparison of the model
predictions and the counts, colors and angular sizes of galaxies in
the HDF. A brief discussion of the results can be found in \S 4, and
the summary and conclusions in \S 5.

\section{Modeling galaxy evolution to fit the deep counts}

In order to get information on galaxy evolution from the deep counts of
galaxies two different approaches have been followed up to now. The
{\it classical} one, pioneered by Tinsley (1972; although she was
trying to understand the Hubble diagram rather than the counts), Kron
(1978), Koo \& Kron (1980) and Shanks (1980), takes as the starting
point the population of galaxies at present (namely, the z=0 Luminosity
Function - LF) tracing back the evolution of the luminosity by assuming
a redshift of formation ($z_{for}$) and a Star Formation Rate (SFR).  A
more sophisticated approach is the one followed by eg. White \& Frenk
(1991), Kauffman, Guiderdoni \& White (1993), Cole et al. (1994) and
Baugh, Frenk \& Cole (1996), in which the starting point is the power
spectrum of primordial density fluctuations predicted by the assumed
theory for structure formation, followed by a {\it recipe} for the
formation of the {\it visible} galaxies in which the dynamics of the
gas, cooling and feedback processes are included.

In this paper I will follow the traditional approach of tracing back
the evolution of the population of galaxies. Even if the
semi-analytical models are undoubtedly proving to give interesting
insight on the problem of galaxy formation and evolution, still the
{\it flexibility} of the simpler approach and the fact that it does not
rely on any specific theory for structure formation can give useful
learning in our interpretation of the deep counts of galaxies.

\subsection{Bright Galaxies}

The population of ``bright'' galaxies has been splitted up into 3 main
types: Ellipticals (E), Spirals (S) and Irregulars (Irr). The z=0 LF
(LF0) has been taken from Efstathiou et al. (1988) with the
morphological mix by Ellis (1983). To compute the evolution of the
galaxy luminosities I used the spectrophotometric models for stellar
population synthesis by Bruzual \& Charlot (1993; new version of
1995). For each type of galaxy there are 3 parameters that we can
adjust, with the constraint that the z=0 model-spectrum has to resemble
the observed spectra of nearby galaxies of the type being
modeled. These parameters are: the redshift of galaxy formation
($z_{for}$), the Star Formation Rate (SFR) and the Initial Mass
Function (IMF). In this work I follow the suggestions given by
Pozzetti, Bruzual \& Zamorani (1996) of using a Scalo IMF for E and S,
because it provides of a ``milder'' luminosity evolution and so the
amount of high-redshift galaxies, observed to be very small in redshift
surveys, is reduced. Similar results can be achieved with the Salpeter
IMF, provided that extinction by dust is also included in the models
(eg. Wang 1991; Franchescini et al. 1994; Koo \& Kron 1995; Campos \&
Shanks 1996). As shown in Table 1, the SFR for E and S is taken to
decay exponentially with time, whereas for Irr I consider a Salpeter
IMF with a constant SFR. For the open models ($q_0=0.05$) the redshift
of formation was taken to be $z_{for}=4$ (Age = 15 Gyr; throughout this
work: $H_0 = 50$ km s$^{-1}$ Mpc$^{-1}$), whereas for the EdS model
$z_{for}=7$ (Age=12.7 Gyr).  The dimming of the luminosity by the Lyman
alpha forest (Madau 1995) and Lyman break is considered in the
modeling of the counts and redshifts. Number density evolution is not
included, even if it might exist at ``moderate'' rates, as already
mentioned in the introduction.

\begin{table*}[h]
\begin{center}
{\sc Table 1: Model Parameters for Bright Galaxies}
\end{center}
\caption{Column 1: Type of galaxy; Columns 2-4: LF0 parameters ($H_0=50$
 km s$^{-1}$ Mpc$^{-1}$); Columns 5-7: IMF, SFR e-folding time and redshift of
formation}
\centering
{\small
\begin{tabular}{c c c c c c c}
 & & & & & & \\
\tableline\tableline
 Type & $\Phi^*$ (Mpc$^{-3}$) & M$^*_B$ & $\alpha$ & IMF &
$\tau_e$ (Gyr) & z$_{for}$ \\
\tableline\tableline
 & & & & & & \\
 E & $9.5\times 10^{-4}$ & -20.9 & -0.48 & Scalo & 1 ($q_0=0.05$) / 0.7 
($q_0=0.5$) & 4 ($q_0=0.05$) / 7 ($q_0=0.5$) \\
 S & $1.15\times 10^{-3}$ & -21.1 & -1.24 & Scalo & 10 ($q_0=0.05$) / 7 
($q_0=0.5$) & 4 ($q_0=0.05$) / 7 ($q_0=0.5$) \\
 Irr & $5.4\times 10^{-4}$ & -21.1 & -1.24 & Salp & constant
& 4 ($q_0=0.05$) / 7 ($q_0=0.5$) \\
\tableline\tableline
\end{tabular}
}
\end{table*}

\subsection{A phenomenological model for dwarfs}

Dwarf (intrinsically faint) galaxies do exist. We see them everywhere
in large numbers. In fact, it has been shown that they may constitute
up to $50\%$ of the whole population in nearby clusters (Binggeli,
Sandage \& Tamman 1985; Ferguson \& Sandage 1991.  See also Trentham
1996). Dwarfs display a whole variety of properties in terms of shapes,
colors and spectral features, while sharing in common the small sizes
and low metal contents. This last property strongly suggests that the
star formation history in these small systems may have followed a
different evolutionary path than in normal galaxies. On this respect an
explanation was first proposed by Dekel \& Silk (1986), who showed that
the shallow potential well in these low-mass galaxies may not be able
to retain the gas after the subsequent galactic wind following an
episode of star formation. Being stripped off the gas, the galaxies
cannot form stars any longer, and so their luminosities, as the massive
stars evolve, fade away. Following this line of arguments, Babul \&
Rees (1992) suggested that the gas may not be entirely lost, but
trapped in the outer parts of the dark halo from where could
re-collapse, given vise to new episodes of star formation. Whether the
gas is completely lost will depend on the mass of the galaxy but also
on the pressure of the environment, that could confine it. This
scenario is in all similar to that first proposed by Davies \&
Phillips (1988), who suggested that the star formation in dwarfs may
proceed in the form of intermittent bursts followed by long quiescent
periods.

The importance of accounting properly for the presence of dwarfs in the
modeling of the deep counts was already shown by Driver et
al. (1994). A much more elaborated model was recently worked out by
Babul \& Ferguson (1996), who very appropriately ``baptized'' the
dwarfs as ``boojums == blue objects observed just undergoing moderate
starburst''. In this work it is claimed that dwarfs may arise in large
numbers in hierarchical models for structure formation at high
redshifts, while the star formation is delayed until $z\sim 1$ due to
the photoionization of the interstellar matter by the UV background
ionization (Babul \& Rees 1992).

Dwarfs are assumed to form continuously since z=1 up to now (see also
Babul \& Ferguson 1996). Because of the ``classical'' approach chosen
to model the deep counts, this continuous formation of dwarfs is
simulated by allowing them to form in contiguous generations each one
following the previous one by 0.5 Gyrs. The LF for dwarfs is assumed to
have a Schechter-like form, with a slope $\alpha=-2$. The reason for
this very steep slope is based on the fact that in hierarchical
clustering the distribution of small halos is a steep function of the
mass.  However, as discussed later, the choice of $\alpha$ won't alter
any of the basic conclusions of this work.

Two different types of dwarfs are tested (called B=0.05 and B=0.5).  In
both of them the star formation is assumed to proceed in the form of a
single burst (i.e. constant star formation) lasting $5\times 10^7$ and
$5\times 10^8$ years for models B=0.05 and B=0.5 respectively. The
masses of dwarfs are the same in both models: $4.7\times 10^{10}
M_{\odot}$ for an $L_*$ dwarf (total mass; I assume that the baryonic
matter is $\sim 1\%$ of the total mass).  The mass of the smallest
dwarfs is $4.7\times 10^8$, or five magnitudes fainter than $M_*$. In
the {\it peak} of star formation this corresponds to a magnitude in B
(assuming a Salpeter IMF) of -18.5 and -20 for B=0.5 and B=0.05
respectively.

The number density of dwarfs is a free parameter chosen to fit the
B-band counts to the faintest limits.  For each generation
$\Phi^*(g(z))=\Phi^*(g(z=0))\times (1+z)^n$, where $g(z)$ refers to the
generation formed at a redshift $z$. Two different cases are tested:
$n=0$ and $n=3$. In the later, dwarfs are assumed to form more
numerously as higher is the redshift of formation.

\begin{table*}[d]
\begin{center}
{\sc Table 2: Model Parameters for Dwarf Galaxies}
\end{center}
\caption{Column 1: Model; Column 2: Duration of the period of star formation;
Column 3: $q_0$; Column 4: rate of dwarf formation (see text); 
Column 5: $\Phi^*$ for the generation of dwarfs forming at z=0; 
Column 6: $M^*$ (in the peak of star formation); 
Column 7: $\alpha$}
\centering
{\small
\begin{tabular}{c c c c c c c}
 & & & & & & \\
\tableline\tableline
 Model & B & $q_0$ & n & $\Phi^*_0 (g(z=0))$ 
(Mpc$^{-3}$) & M$^*_B$ & $\alpha$ \\
\tableline\tableline
 1 & 0.05 & 0.05 & 0 & $5\times 10^{-3}$ & -20 & -2 \\
 2 & 0.05 & 0.5 & 0 & $5\times 10^{-3}$ & -20 & -2 \\
 3 & 0.5 & 0.05 & 0 & $3\times 10^{-3}$ & -18.5 & -2 \\
 4 & 0.5 & 0.5 & 0 & $3\times 10^{-3}$ & -18.5 & -2 \\
 5 & 0.5 & 0.5 & 3 & $1\times 10^{-3}$ & -18.5 & -2 \\
\tableline\tableline
\end{tabular}
}
\end{table*}

\section{Dwarfs with gentle star formation and the deep counts}

\subsection{HST counts as a function of morphology}

The unprecedented high-resolution imaging capability of the Hubble
Space Telescope makes now possible to study the shape of very distant
(faint) galaxies. Deep counts {\it as a function of morphological type}
have recently been published (Driver et al. 1995; Glazebrook et
al. 1995a; Abraham et al.  1996) down to I=25. It has been shown that
the counts of E galaxies increase much more slowly than the counts of
irregular/peculiar systems (also called ``weirdos'' galaxies;
hereafter W).  In fact, the counts of E show a small evidence for
flattening at I$\sim 25$. At I=18 E are more numerous than W by a
factor of $\sim 3$, while at I=25 W become more numerous than E by
almost a factor of 2, reaching a level of counts comparable to that of
S. These results have led to questioning the extent to which the Hubble
system provides an adequate description of the morphology of galaxies
at high redshifts.

The I-band counts splitted into the 3 morphological types (E, S and W)
are shown in Figure 1 together with the model predictions. As it can be
seen in the Figure, the {\it pure luminosity evolution models} (PLE)
provide reasonable predictions for the counts of E and S, while
severely under-estimate the number of W unless a (numerous) population
of dwarfs is included (see Table 2 for more details on the dwarf models
shown).

\begin{figure}
\centering
\centerline{\epsfysize=9.5truecm
\figinsert{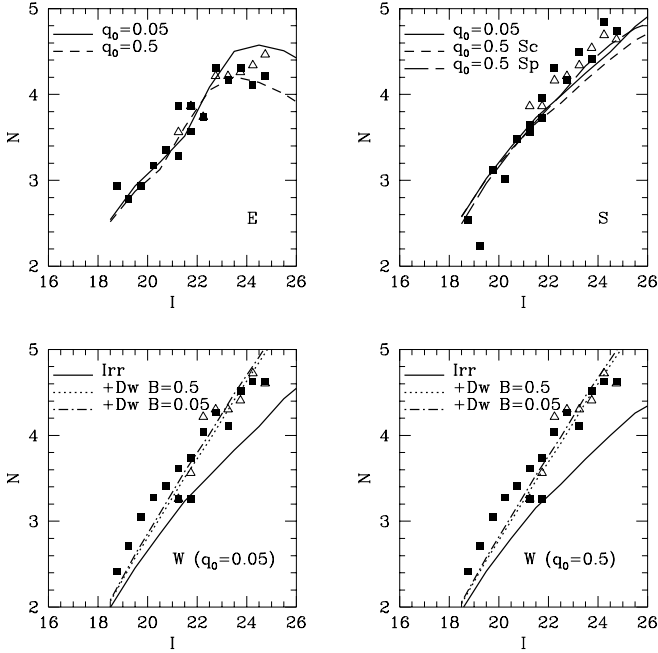}{Figure 1}}
\caption{The counts of galaxies as a function of
morphological type (data kindly provided by R. Abraham) in the I
(Kron-Cousins) band.  Together with the data there are shown pure
luminosity evolution models with and without a population of dwarfs to
fit the counts of ``weirdos'' (W) galaxies. In top panels there are
shown counts of E and S galaxies. Solid and dashed lines are for
$q_0=0.05$ and $q_0=0.5$ PLE model predictions respectively. For S both
a Salpeter IMF (short dashed line) and a Scalo one (long dashed line)
were considered. In bottom panels the counts of W are plotted, together
with model predictions ($q_0=0.05$ - left panel and $q_0=0.5$ - right
panel). Solid lines are the predictions when only a population of
Irregulars is considered, whereas dashed lines are predictions from
models including a population of dwarfs with a burst length of B=0.5
Gyr (dot lines) and B=0.05 Gyr (dot-dashed lines).  For details on the
modeling see Tables 1 and 2 and text.}  
\end{figure}

\subsection{B-I color distribution}

The fact that PLE models may have problems to predict the level of
counts at very faint magnitudes is further evidenced in Figure 2. In
this Figure it is shown the distribution of B-I colors for the
galaxies in the Hubble Deep Field (Metcalfe et al. 1996), for galaxies
selected according to their magnitudes in the I-band (top panels) and
in the B-band (bottom panels).  To put the HDF counts into the
standard Johnson system in order to compare the HDF results with
ground-based data (as in Figures 6a-d, where deep counts from a variety
of sources are shown), it is necessary to use certain color
conversions.  The galaxy counts shown here were worked out by Metcalfe
et al. (1996; 1997) who used the synthetic color transforms of
Holtzman et al. (1995) and the published values of the HDF zero-point
conversion from STMAG to Vega system. The details of the procedure can
be found in Metcalfe et al. (1997). As shown in Metcalfe et al. (1996),
there is an excellent agreement between space and ground-based data
down to the faintest limits (e.g. $B\sim 28.2$, the limiting magnitude
of the ``Herschel Deep Field'').

\begin{figure}
\centering
\centerline{\epsfysize=9.5truecm
\figinsert{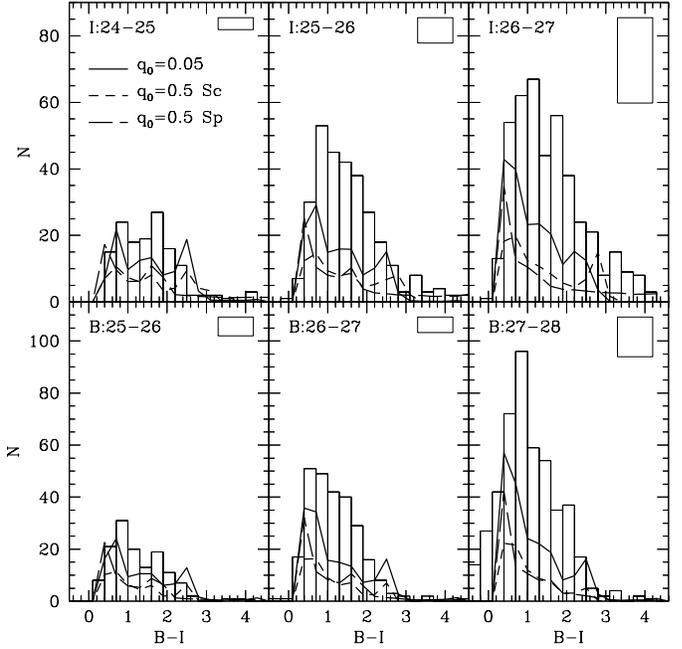}{Figure 2}}
\caption{The B-I color distribution of faint galaxies in
the HDF selected in the I-band (top panels) and B-band (Johnson;
bottom panes) for different magnitude bins. The boxes in each panel
corresponds to the incompleteness. Also there are shown predictions
from pure luminosity evolution models. The models are not normalized to
the number of galaxies in each magnitude bin but corresponds to the
expected number of galaxies to be seen in the HDF field of view for
each model.}  
\end{figure}

Together with the data, they are plotted predictions from PLE
models. It is important to notice that the models {\it are not
normalized to the number of galaxies in the HDF for each magnitude
bin}, but just correspond to the predicted number of galaxies to be
seen in the HDF field of view. The normalization is fixed by the
assumed values of $\Phi^*$ (see Table 1), which already corresponds to
a \it high \rm normalization, B$\sim 18$.  To the eye, the color
distribution from PLE models seems to be a good match to the data if
models had been normalized to the total number of galaxies in each
magnitude bin. However if we did so, the deep counts would be
over-predicted up to very faint levels. For example, in the $q_0=0.05$
model, data and predictions almost agree for $24<I<25$, but to fit the
data for fainter bins with the same model we would have to multiply the
normalization of the model by almost a factor of 3. Therefore the
result which emerges is that, even if the predicted color distribution
is similar to the observed one, a match to the data requires a
normalization inconsistent with the counts. A remaining question is
that of the very red high-z Lyman break galaxies. For the open model,
because $z_{for}=4$ there are no Lyman-break galaxies to be detected in
the B-band (i.e. the break enters the B-band at $z\sim 4$, and so these
galaxies would show up as very red in $B-I$ if they were located at
$z>4$). In the EdS models $z_{for}=7$, and so the situation is
different. However, because the volume element for $z>4$ when
$q_0=0.5$ is very small, the contribution of these galaxies is
negligible.

As said before, none of the models is able to provide a reasonable fit
when using the appropriate normalization.  Interestingly, as we approach
fainter limits, the models more severely under-predict the number of
{\it red} faint galaxies (i.e. B-I$\sim 1-2$) observed.

The same plot is shown in Figures 3 and 4 (for $q_0=0.05$ and 0.5
respectively), although now the models include the dwarf population
(n=0 case).  As expected, the level of counts increases and the color
distribution is much better reproduced.  It becomes however clear that
the star formation rate assumed for the dwarfs is a ``key'' issue to
reproduce the data. The B=0.5 model giving a much better fit to the
color distribution than the B=0.05 one.

\begin{figure}
\centering
\centerline{\epsfysize=9.5truecm
\figinsert{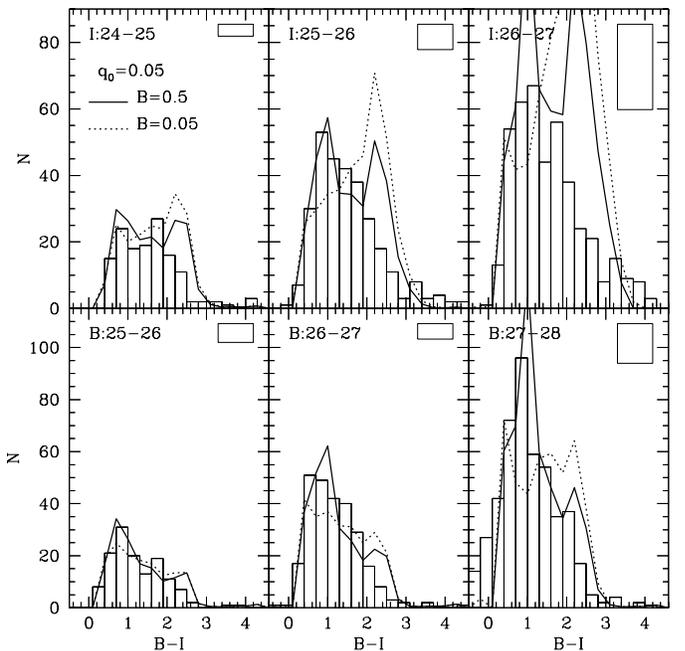}{Figure 3}}
\caption{The same than Figure 3, but models now incorporate
a population of dwarf galaxies with the star formation lasting 0.05 and
0.5 Gyr ($q_0=0.05$).}  
\end{figure}

\begin{figure}
\centering
\centerline{\epsfysize=9.5truecm
\figinsert{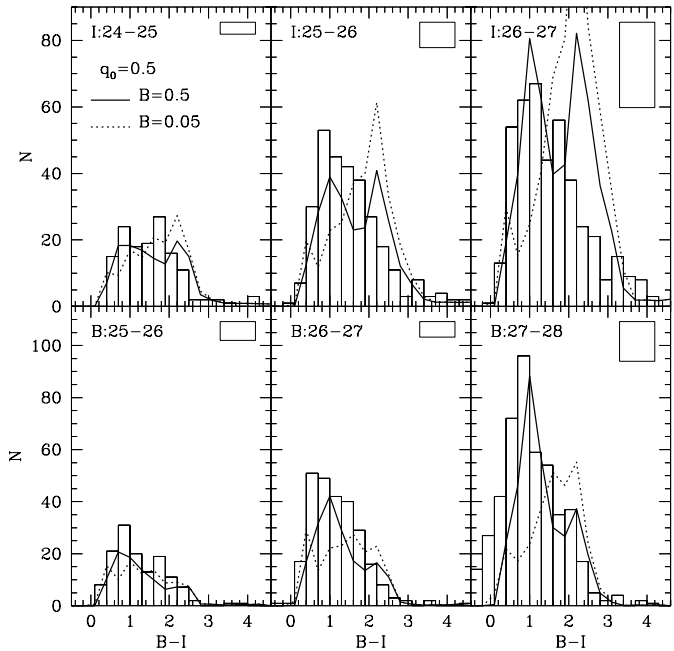}{Figure 4}}
\caption{The same than Figure 3, but models now incorporate
a population of dwarf galaxies with the star formation lasting 0.05 and
0.5 Gyr ($q_0=0.5$).}  
\end{figure}

As shown in Table 2, the number density of dwarfs in the B=0.05 model
is larger than in the B=0.5 one for almost a factor of 2. The reason is
found in the restriction imposed that the model has to be able to
(approximately) reach the observed level of counts in the B-band.  In
the B=0.05 model, because the galaxies form the bulk of stars during a
shorter period, fading away afterwards, the number of dwarfs required
to fit the B-band counts is larger.  This is simply due to the fact
that the probability of observing a B=0.05 dwarf while exhibiting \it
blue \rm colors (i.e. in the ``boojum'' phase) is smaller than for a
B=0.5 one.  As a result, the B=0.05 model predicts a large population
of red remnants which is not seen in the color distribution of the HDF
galaxies. The disagreement is even worse if $q_0=0.5$.

The fit to the color distribution provided by the B=0.5 model is quite
reasonable. It is only for the I=26-27 magnitude bin where model and
data show disagreement, the model predicting a larger number of red
(B-I$\sim 2-3$) galaxies than it is observed.  However for this
magnitude bin the incompleteness is large (see box in the Figures), and
the galaxies without measured color (i.e. those detected in the I-band
image but not in the B-band one) are expected to be red, i.e. too faint
(B$\sim 29-30?$) to be detected.

It is interesting to notice that the shorter the star formation period
in dwarfs (notice that Babul \& Ferguson use $\sim 10^7$ yrs) the
larger the number of (un-observed) remnants and vice-versa. Therefore
if we want to fit the counts to the faintest levels by means of dwarfs,
the star formation has to proceed in a more ``gentle'' way.

In order to test the effect of the rate at which dwarfs are being
formed, I show in Figure 5 the same plot with predictions from B=0.5
models with two different rates of dwarf formation: n=0 (formation of
dwarfs constant with time) and n=3 (formation of dwarfs decreases with
time). The differences between the two are not as large as between the
B=0.05 and B=0.5 models. Still it can be seen that in the n=3 case the
number of red remnants in the I=26-27 bin is larger, as it is
expected. Nevertheless the data does not allow a clear distinction
between the two models.

\begin{figure}
\centering
\centerline{\epsfysize=9.5truecm
\figinsert{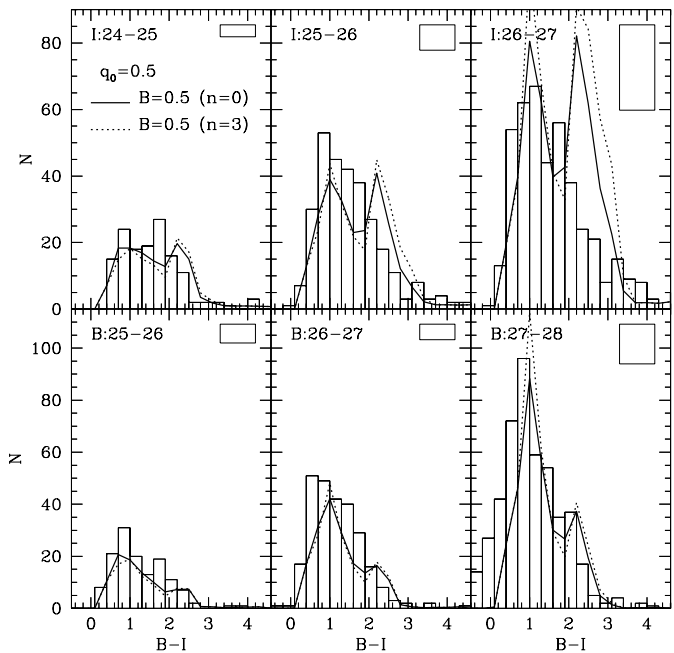}{Figure 5}}
\caption{The same than Figure 3, but models now incorporate
a population of dwarf galaxies with the star formation lasting 0.5
Gyr. The rate of dwarf formation with redshift is assumed to be
$\Phi^*(g(z))=\Phi^*(g(z=0))\times (1+z)^n$. Here two models are
tested: n=0 and n=3 ($q_0=0.5$).}  
\end{figure}

\subsection{Deep counts and redshift distributions}

The deep counts in the K- (data taken from Djorgovski et al. 1995;
Gardner et al. 1996a, 1996b; Soifer et al. 1994; McLeod et al. 1995;
Glazebrook et al. 1995), I- (Metcalfe, Shanks \& Fong 1995, Metcalfe et
al.  1996a; Driver et al. 1994, 1995; Glazebrook et al. 1995; Smail et
al. 1996; Lilly, Cowie \& Gerdner 1991; Tyson 1988; Hall \& Mackay
1984; Koo 1986), B- (Metcalfe et al. 1991, 1995, 1996a; Lilly et
al. 1991; Tyson 1988; Couch \& Newell 1984; Infante, Prittchet \&
Quintana 1986; Jones et al. 1991; Koo 1986; Kron 1987; Maddox et
al. 1990) and U-band (Jones et al. 1991; Metcalfe et al. 1996;
Guhathakurta, Tyson \& Majeswki 1990; Koo 1986) are shown in Figure
6a-d together with various model predictions for B=0.5 and 0.05 and
$q_0=0.05$ and 0.5.  (Because the counts are plotted in a
logarithmic scale, they have been normalized by subtracting the
corresponding ``best'' fit at brighter levels in order to expand the
scale).  It must be noticed here that the HST F300W filter is not very
close to the standard U band, what complicate any comparison between
HDF and ground-based data (see Metcalfe et al. 1997).  Because the main
source of data at the very faint limits is the HDF, the comparison
between models and data in the U basspand must be taken with caution
not only because of the uncertainties in the color conversion but
because it can affect the number of Lyman dropouts as the F300W
wavelength is shorter than the standard U-band one. 

In the K-band all 5 models give good predictions, although it would
seem that the EdS case fits the data slightly better.  In particular
the shoulder (clear change of slope) seen at K$\sim 10$ is very nicely
reproduced. Notice that there is very little difference between the
B=0.5 and B=0.05 models, the reason being that the contribution of the
dwarfs to the K-band counts is only noticeable at very faint levels.
In the other bands all 5 models are marginally consistent with the
data, although the B=0.5 case give more consistent fits. For example,
the B=0.05 $q_0=0.5$ model clearly over-predicts the I-band counts
while under-predicting in the U-band.  (Notice that the number of
Lyman dropouts could be under-predicted because, as pointed out above,
the F300W wavelength is shorter than the standard U-band one used in
the models. This would affect the modeling in the sense of
``over-predicting'' the counts.) It should be noticed that the I-band
counts are better fitted by EdS models than by open ones.  Faintwards
than $I\sim 26$ open models over-predict the counts, especially when
B=0.05. 

\begin{figure}
\centering
\centerline{\epsfysize=9.5truecm
\figinsert{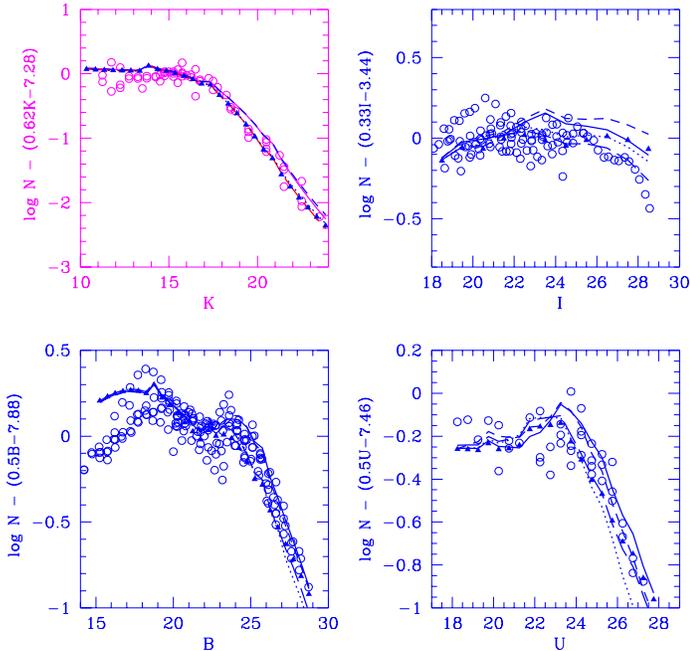}{Figure 6}}
\caption{The deep counts of galaxies (per 0.5 mag per sq deg) 
in the K-, I-, B- and
U-band (open circles; data taken from the literature. See text for
references). Predictions from different luminosity evolution models
with dwarfs are shown: triangles - ($q_0=0.5$ B=0.5 n=3); solid lines -
($q_0=0.05$ B=0.5 n=0); short dashed lines - ($q_0=0.05$ B=0.05 n=0);
long dashed lines - ($q_0=0.5$ B=0.5 n=0); dot lines - ($q_0=0.5$
B=0.05 n=0).}  
\end{figure}

A further test to the models can be done by means of the redshift
distribution of galaxies n(z) selected in different photometric
bands. In fact the very early PLE models (eg. Metcalfe et al. 1991)
that were successful in fitting the deep counts faced some problems with
the absence of a {\it high redshift tail} in the observed n(z). These
first discrepancies were solved by including further ingredients in the
modeling, like the presence of dust (eg. Wang 1991; Franchescini et
al. 1993; Gronwall \& Koo 1995; Campos \& Shanks 1996) or variations in
the IMF (Pozzetti et al. 1996). Also the introduction of merging
(Guiderdoni \& Rocca-Volmerange 1991; Broadhurst et al. 1992) helped to
solve the problem.

Here I'll use the sample of Colless et al. (1991; 1993) for galaxies
with B:21-22.5, the recent survey by Cowie, Songalia \& Hu (1996) for
galaxies with B:22.5-24 and K:18-19, and the Canada-France Redshift
Survey (CFRS; Crampton et al. 1995) for galaxies in the magnitude range
I=17-22. The n(z) for the 4 samples are shown in Figure 7 together with
model predictions. As can be seen, the n=0 models face some problems as
slightly over-predict the number of low-z galaxies in Colless et
al. (which is an almost ``complete'' sample). The number of very low-z
galaxies is severely reduced when $n>0$ (i.e. if the formation of
dwarfs decreases with time). It is in any case interesting to point out
that it is usually claimed (eg. Glazebrook et al. 1995b) that the
un-identified galaxies in deep redshifts surveys are likely to be at
high-z. The reason given being that the redshifts could not be measured
because the main spectral features (like the frequently strong
[OII]$\lambda 3727$ emission line) are red-shifted outside the optical
spectral window. However it may happen that some ``still-blue and
luminous'' dwarfs show a featureless spectrum, and so their redshifts
are difficult to be measured. To illustrate this it is shown in Figure
8 the evolution with time of the [OII]$\lambda 3727$ equivalent width
and B-I color after a single burst of star formation. As soon as the
massive stars evolve (which happens very fast) the equivalent width
drops to zero, as there are no UV photons capable of ionizing the
interstellar medium.  While the colors remain blue for quite a longer
period of time due to the bulk of intermediate mass stars still in the
main sequence. Therefore, some of the un-identified objects might be
dwarfs at low-z, still blue and luminous but in which the star
formation just ceased.

\begin{figure}
\centering
\centerline{\epsfysize=9.5truecm
\figinsert{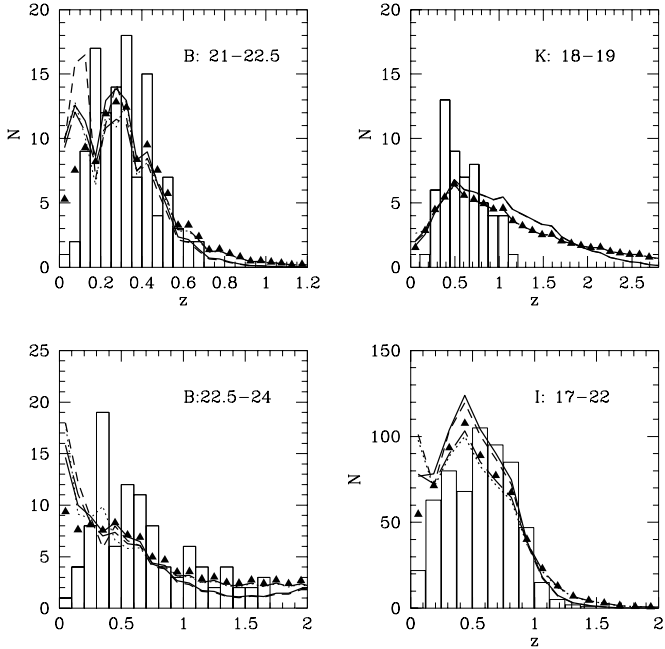}{Figure 7}}
\caption{The n(z) redshift distribution of galaxies
selected in different magnitude ranges (data taken from the literature.
See text for references). The incompleteness rate (i.e.  galaxies for
which the redshift could not be measured) are: $\sim 12\%$ for
B:21-22.5, $\sim 15\%$ for I:17-22 and $\sim 20\%$ for K:18-19 and
B:22.5-24.  Lines are predictions from n=0 dwarf models as in Figure
6. Notice that B=0.5 models predicts less amount of very low-z dwarfs
than B=0.05. It is the n=3 model (B=0.5, $q_0=0.5$; triangles) which
provides the best fits to all the data. The percentage of galaxies
predicted by the models to be located beyond the highest redshifts
shown in the Figure are: $\sim 4\%$ ($q_0=0.5$) and $\sim 1\%$
($q_0=0.05$) for B:21-22.5, $\sim 10\%$ ($q_0=0.5$) and $\sim 2\%$
($q_0=0.05$) for I:17-22, $\sim 6\%$ ($q_0=0.5$) and $\sim 0.4\%$
($q_0=0.05$) for K:18-19 and $\sim 17\%$ ($q_0=0.5$) and $\sim 24\%$
($q_0=0.05$) for B:22.5-24. The three flat models on the one hand, and
the two open models on the other, give very similar percentages for the
predicted high redshift population.}  
\end{figure}

\begin{figure}
\centering
\centerline{\epsfysize=9.5truecm
\figinsert{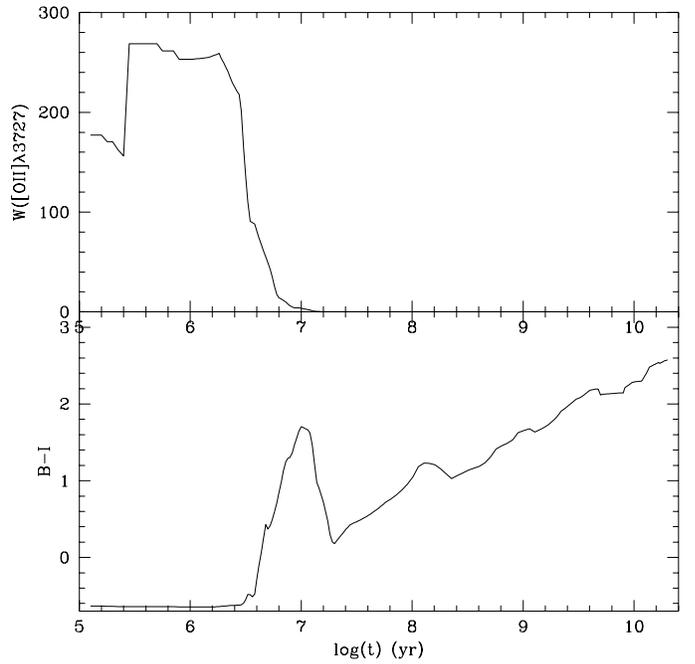}{Figure 8}}
\caption{The evolution of the [OII]$\lambda$ 3727
equivalent width and the B-I color with time after a single,
instantaneous burst of star formation (the model has been kindly
provided by G. Magris).}  
\end{figure}

 In order to give the reader a better view on the behavior of
the models, in Figure 9 they are plotted predictions for the redshift
distribution of galaxies at fainter magnitude bins (i.e. $24<B<25$,
$25<B<26$, $26<B<27$ and $27<B<28$). As expected for dwarf-rich models,
at the very faint limits most galaxies are located at low redshifts
(i.e. $z<1$). There are however differences from model to model. For
example, it can be seen that the B=0.05 models predict a higher
percentage of galaxies at $z<0.2$ than the B=0.5 ones, especially in
the faintest bin. Also, the distribution is shifted toward higher
redshifts when n=3 instead of n=0.

\begin{figure}
\centering
\centerline{\epsfysize=9.5truecm
\figinsert{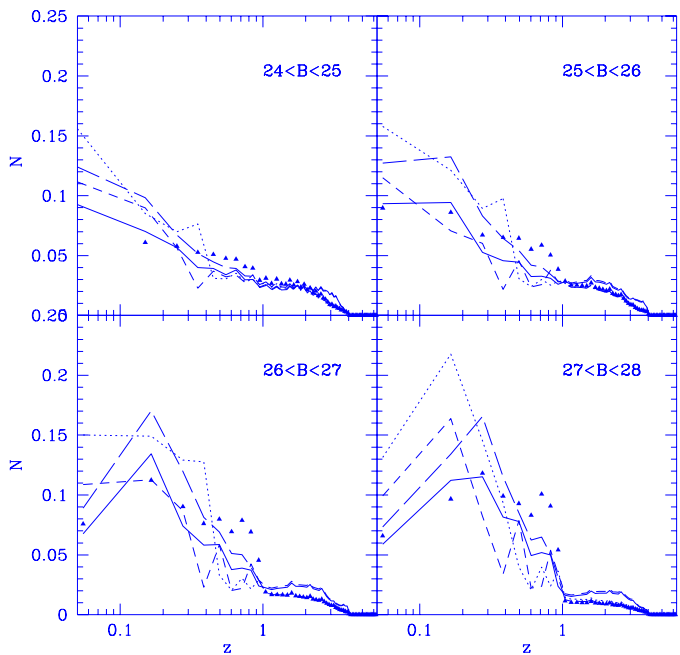}{Figure 9}}
\caption{Model predictions for four (faint) magnitude
bins. Lines and triangles are for different dwarf models as in Figure
7.}  
\end{figure}

As a further test to the models, in Figure 10 it is plotted the z=0 LF
used, together with the LF measured by Loveday et al. (1992) and Marzke
et al. (1994) (arbitrarily normalized, because of the inconsistency
between the two LFs). The LF0 shown here correspond to the B=0.5
model. It can be seen that the steep slope measured by Marzke et
al. (1994) is, within the errors, consistent with the model. Notice
that the slope of the LF for each generation of dwarfs is very steep,
but the superposition of all dwarfs from the different generations plus
giants at z=0 gives a much flatter slope, except at magnitudes fainter
than $\sim -15$.

\begin{figure}
\centering
\centerline{\epsfysize=9.5truecm
\figinsert{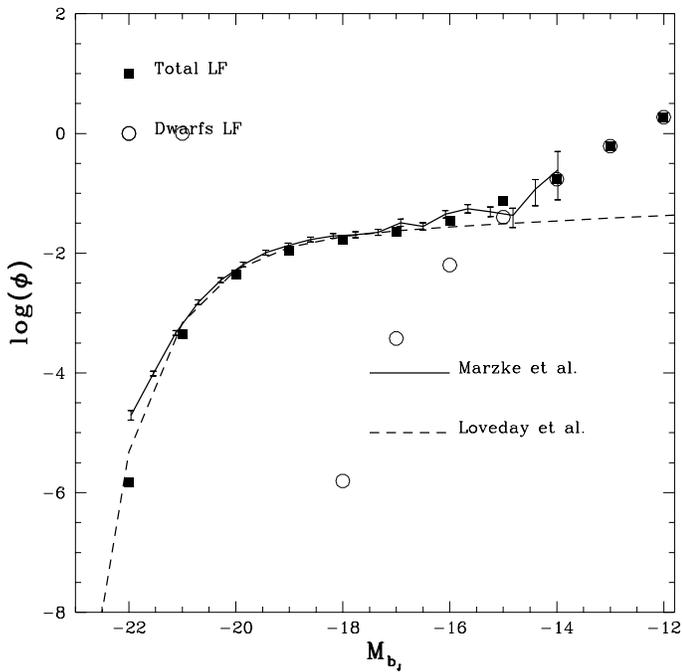}{Figure 10}}
\caption{The z=0 LF from Loveday et al. (1992) and Marzke
et al.  (1992; the magnitudes has been shifted by 0.7 to fit Loveday's
LF). Also shown is the LF0 in the B=0.5 n=0 model.}  
\end{figure}

\subsection{Angular size distribution}

\begin{figure}
\centering
\centerline{\epsfysize=9.5truecm
\figinsert{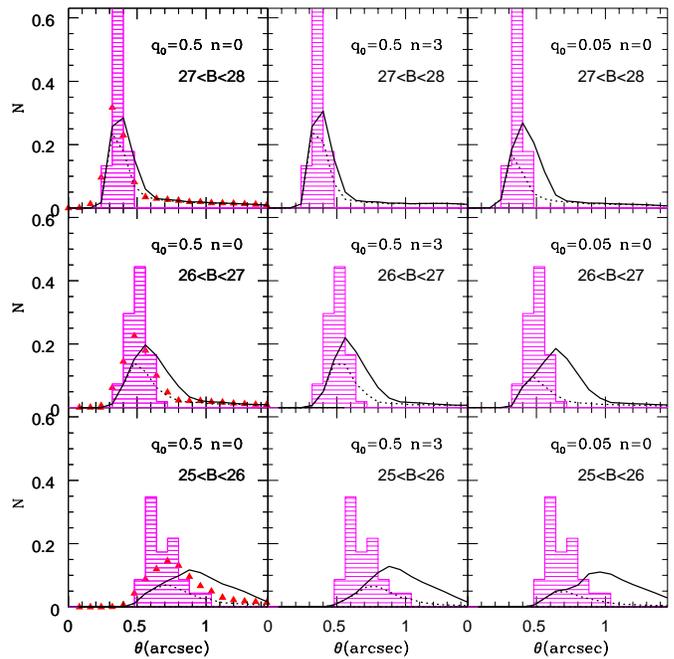}{Figure 11}}
\caption{The isophotal angular size distribution of HDF
galaxies (see text) and predictions from different B=0.5 models. Solid
lines are for the total population of galaxies, whereas dotted lines
are for the dwarf galaxies. Triangles are predictions from the model
when spiral disk evolution is included.}  
\end{figure}

One of the most ``puzzling'' results that came out from the HDF
concerns the sizes of the galaxies. The HDF is filled with apparently
too many ``little tiny'' galaxies. Merging may help to solve the
problem, by assuming that galaxies at high redshifts are splitted into
the (smaller) fragments that will eventually merge to build up the
present-day population (see Im et al. 1995, Roche et al. 1997). Another
possible explanation is found in the dwarf-rich models, where a large
contribution to the faint counts comes from intrinsically small (dwarf)
galaxies. In order to check the viability of the models analised here
next I show predictions for the angular size distribution of the
galaxies in the HDF. (The isophotal angular sizes of the HDF galaxies
shown in the Figures were measured by N. Metcalfe, who kindly provided
me with the data previous to publication).

To compute the angular size distribution it is first neccesary to find
relationships between physical size and absolute z=0 B-band luminosity
of the galaxies. As in Im et al (1995), for ellipticals the scaling
laws derived by Binggeli, Sandage \& Tarenghi (1984) were used;
$\log(r_{hl}/kpc)=-0.3(M_B+18.75)$ if $M_B<-20$ and
$\log(r_{hl}/kpc)=-0.1(M_B+15.7)$ if $M_B>-20$, where $r_{hl}$ is the
model-independent half-light radius. The profiles are assumed to fit a
de Vaucouleurs law, and the Kormendy's relationship is used to relate
the effective radius with the effective surface brightness
($\mu_{eff}$): $\mu_{eff}=-2.5\log(r_{eff})+20.2$ (the zero-point of
the relation was taken from the work by J\o rgensen, Franx \& Kj\ae
rgaard 1995 for ellipticals in the Coma cluster). Notice that here I
assume the half-light radius to be the same than the de Vaucouleurs
effective radius, which will only be true as long as the de Vaucouleurs
law is a good fit to the light profiles (see Binggeli, et
al. 1985). The angular sizes of the HDF shown in the Figures are
isophotal sizes, defined by the isophote $\mu_B=28.5$ mag per square
arcsec.  Assuming a de Vaucouleurs profile, the physical $r_{28.5}$
size corresponding to the $\mu_B=28.5$ isophote is computed by taking
into account the surface brightness dimming:
$\mu^*_{28.5}=28.5-10\log(1+z)$. $\mu^*_{28.5}$ is the surface
brightness of the (z=0) isophote that, at a redshift $z$, would be
observed as 28.5 mag per arcsec square (see e.g. Sandage 1961). Then,
$r_{28.5}=r_{eff}[(\mu^*_{28.5}-\mu_{eff})/8.325+1]^4$. Notice that the
evolution of the luminosity (and the K-correction) is included in the
modeling, but no size evolution has been assumed (this means, a
uniform fade of the luminosity is considered). Regarding the modeling
of ellipticals, it should be quoted here the work by Im, Griffiths \&
Ratnatunga (1997) who tested luminosity evolution and merging models by
means of the distribution of angular sizes and colors in the Hubble
Space Telescope Medium Deep Survey. The modeling of the ellipticals
follow the same approach as in Im et al. (1995), which, as quoted
before, is also followed in the present work. Im et al. (1997) show
that for $20<I<21$ and $21<I<22$ simple luminosity evolution models
(taken $q_0=0.5$ and $H_0=50$ km s$^{-1}$ Mpc$^{-1}$) nicely reproduced
the half-light angular size distribution whereas merging models predict
too many small galaxies compared to the data.

According to Freeman's law (Freeman 1970), the central surface
brightness of the disk component in spiral galaxies is nearly constant
($\mu_0(B)\sim 21.3$). Even if not properly understood yet, results
have accumulated indicating the validity of this empirical law (Boroson
1981; Kent 1985; Bosma \& Freeman 1993).  The fact that the central
surface brightness is constant (though with a scatter of about 0.5-1
mag., see e.g. de Jong 1995, Marquez \& Moles 1997) implies that the
scale length and the luminosity of disk components are correlated:
$\log(r_0/kpc)=-0.2M_B-3.45$. As pointed out by Im et al. (1995), in
modeling the spirals complications arise because the galaxies are not
pure disks but also show a bulge component. To overcome this problem,
Roche et al. (1997) model the half-light radii of spirals $r_{hl}$ by
considering that $r_{hl}=1.4r_0$ for the early-types, instead of the
``pure-disk'' relationship $r_{hl}=1.68r_0$ which is used for the late
types.  In the present work, because a single spiral type is
considered, all spirals have been modeled as pure disks. Therefore we
will have to keep in mind that the sizes will be slightly
over-estimated. As for the ellipticals, the surface brightness dimming
is taking into account in computing the isophotal angular sizes, and no
size evolution is considered. I'll get back to this point at the end of
the section.

The above empirical relationships allow, in a first simple approach, to
compute predictions for the angular size distribution of ``bright''
galaxies. For dwarfs the situation is much more uncertain. As pointed
out by Impey, Bothun \& Malin (1988), studies about the structural
parameters of dwarf galaxies are ``obscured'' by all kind of selection
effects due to the intrinsic faintness of the objects. Nevertheless it
is already well known that the light profiles of dwarf ellipticals (dEs)
obeys a simple exponential law (Binggeli et al. 1984; Ichikawa,
Wakamatsu \& Okamura 1986; Caldwell \& Bothun 1987; Impey et al.
1988).  However for the dwarf irregulars (dIrrs) the situation is less
clear as some of them show an excess of light over the exponential fit
in the central parts (Bothun et al. 1986). With respect to the Blue
Compact Dwarfs (BCDs), at least some of them seem to have surface
brightness profiles that can be fitted by exponentials (Bothun et
al. 1986; but see also Kunth et al. 1988). This situation has led to
the suggestion that (some of) the BCDs might be the truly progenitors
of the dEs, but it is less clear that dIrrs could evolve into dEs. On
this respect Bothun et al. propose that there might be a continuous
spectrum of dwarfs, and even if they all could eventually be stripped
off the gas, the gas depletion process as well as the star formation
efficiency might be quite complex, probably related to the
gravitational potential well in which the gas in embedded.

Compared to the picture that emerges from the observations, the
modeling of dwarfs in the present work is obviously quite
simplistic. A single type of dwarf is considered, with a star formation
process lasting 0.05 and 0.5 Gyr (depending upon the model
considered). These numbers must be taken as a very first approximation
to the star formation period of the ``average'' dwarf, as it seems
clear that the star formation process and the gas depletion will not
be the same for all dwarfs. Nevertheless it should be pointed out that,
by allowing the star formation process to last over a relatively long
period of time, the model is implicitly assuming that the gas
depletion is more complex than in the simpler case in which the
galaxies fade away after a single burst of star formation.  A second
problem to deal with is the evolution of the galaxy sizes as the
luminosities fade. On this respect Binggeli (1985) has proposed two
different scenarios: 1) uniform fade, i.e. the scale length will be
preserved, and 2) faster fade in the inside regions where the star
formation could be more concentrated.

Continuing with the simple approach in the modeling of dwarfs, here
I'll assume a single exponential fit for all dwarfs with constant
central surface brightness \it at the peak of the star formation
process, \rm ($\mu_0(peak)\sim 21$), together with a uniform fade in
luminosity. Like for the spirals, an exponential profile with a
constant central surface brightness (that of course will fade at the
same ratio than the luminosity) implies a correlation between total
luminosity \it at the peak of star formation \rm and scale length.  For
the B=0.5 model, where $M_{peak}^*=-18.5$ (see Table 2), a $M^*$ dwarf
would have a disk scale length of $\sim 3.5$ kpc, whereas for the
smallest dwarfs ($M_{peak}=-13.5$) $r_0\sim 0.4$ kpc.  The existence of
a correlation between luminosity and scale length for dwarf ellipticals
has been shown by Binggeli et al. (1985) among others.  For BCDs,
Campos, Moles \& Masegosa (1990) have shown the existence of a
well-defined correlation between luminosity and isophotal radius of the
form: $\log(L_{25}(B)) \propto 1.88\log(R_{25})$. The slope of the
correlation, very close to 2, implies a nearly constant surface
brightness inside the 25 mag per square arsec isophote, as it is in
fact observed ($\mu_{25}\sim 23.5$) though with a large ($\sim 1.5$
mag) scatter. The scatter is interpreted in this paper as reflecting
the fade in luminosity already at the BCD stage. Assuming an
exponential surface brightness profile for dwarfs, and taking
$\mu_{25}\sim 23.5$ for BCDs which could be considered as dwarfs
observed $\sim$ very close to the \it peak of the star formation
process, \rm it is derived that $\mu_0(peak)\sim 21$.

In Figure 11 they are shown predictions for the B=0.5 model for 3
different cases: $q_0=0.5$ (with $n=0$ and $n=3$), and $q_0=0.05$ (with
$n=0$).  Solid lines are predictions for the total angular size
distribution, whereas dotted lines correspond to the dwarfs. The
histograms are normalized to the total number of galaxies in each of
the magnitude bins. As can be seen, for the faintest bin ($27<B<28$)
the model(s), in spite of its simplicity, fits remarkably well the data
(especially if $q_0=0.5$). For the brighter bins, in particular for the
$25<B<26$ one, the models predict too large sizes compared to the
data. It should be noticed that the contribution to the ``un-observed
large'' galaxies comes from the ``bright'' population, as dwarf sizes
fall well into the observed range.

In order to get some insight into the sources of discrepancy between
data and models, (even if this goes beyond the goals of this paper
which mainly addresses the role of dwarfs in the faint counts), let us
go back to the modeling of the spirals. As said before, spirals are
treated as single ``pure'' disks with no size evolution. Already the
first assumption, i.e. neglecting the bulge, makes the sizes to be
over-estimated. Also, it has been largely suggested in most models
about the formation and evolution of disk components (see e.g. Lacey \&
Fall 1985; Wang \& Silk 1994; Cayon, Silk \& Charlot 1996), that disks
are very likely formed from inside outwards as the gas is slowly
falling from the halo to develop the disk . Just to check the effect of
the ``disk growth'' in the predictions, let us model it in a very crude
way as: $\log(r_0/kpc,z)=-0.2M_B+3.45(1+n\times z)$ with
$n=0.03$. (This means, for $z=1$ and $z=2$ galaxies the disk lengths
will be a factor of $\sim 1.2$ and $\sim 1.5$ smaller than for $z=0$
galaxies). Adding this (totally) ``ad hoc'' disk evolution, the
predictions now (triangles in the Figure) fit the data also for the
brighter bins and discrepancies smooth away.

\section{Discussion}

The phenomenological model for dwarfs shown here is somewhat {\it ad
hoc}, due to the lack of knowledge about how dwarfs are formed are
evolved.  Nevertheless it is based on reasonable assumptions such as
the formation at low-z when the UV-background ionization decreases
(Babul \& Rees 1992) or stars being formed during a ``finite'' period,
the luminosity of the galaxy fading away afterwards (if stars where
continuously formed, the metallicity would not keep at the low levels
observed among dwarfs).

For how long does the star formation period take place? If counts and
colors in the HDF are to be fitted by means of the dwarf population,
the star formation period has to be quite long, at least few times
$10^8$ years.  As it was shown in the last section, using shorter
periods requires of larger numbers of dwarfs to fit the deep counts,
which results in a large number of red, faint remnants not seen in the
HDF.  (It should be said here that, for the B=0.05 model, a
time-consuming trial and error test - i.e. variations of $\alpha$, n,
$M_*(peak)$ - showed that there is no choice of a set of parameters
able to provide good fits to the data).

In the ``standard'' model for dwarfs it is suggested that the star
formation takes place in the form of single, very short bursts. As
first shown by Dekel \& Silk (1986), the shallow potential well of the
galaxy is not likely to be able to retain the gas after the explosions
of few supernovae. This vision has been challenged by the analysis of
the photometry of individual stars in nearby dwarfs. For example, in
the Carina dwarf spheroidal galaxy by Smecker-Hane et al. (1996). In
this very low mass galaxy the star formation history has been very
complex: several bursts each one lasting $\sim 1$ Gyr followed by
quiescent periods. In order to explain this complex history together
with the low metal content in the Carina galaxy, the authors suggest
that the ``gentle'' star formation during the ``active'' phases
generated winds which expelled the metal-enriched gas. But in order to
keep the star formation without stripping the galaxy off, these winds
cannot be strongly coupled with the general interstellar medium, and so
denser gas clouds are not expelled from the galaxy. Therefore there
must exist a sort of self-regulation mechanism able to keep a low-rate
(gentle) star formation process for long periods without increasing the
metal content. The gas might be eventually lost, but the gas depletion
could be a very slow process, perhaps due to the presence of a massive
dark halo.

A case of {\it anomalous} chemical enrichment in local dwarfs was found
in GR8. This low metallicity galaxy ($Z \sim 1/18 Z_{\odot}$) shows a
high helium content, a fact interpreted as the result of a selective
metal lost during a {\it gentle} process of star formation (Moles,
Aparicio \& Masegosa 1990). In the same line, Masegosa, Moles \& Campos
(1994) analised a sample of 121 HII galaxies, finding no trend between
the Helium abundance and the metallicity of the systems. Again, this
result was interpret as reflecting the inability of the galaxy to
retain the supernova ejecta, though the star formation phase in those
systems is still an on-going process.

In the best model (B=0.5) for dwarfs used here each galaxy under-goes a
single period of star formation lasting 0.5 Gyr, fading away
afterwards.  If the period of activity was longer and/or galaxies
under-went multiple bursts, the number of dwarfs requires to fit the
deep counts would be smaller, possibly in a better agreement with the
observed n(z) distributions.

The B=0.5 model should be taken as a simple ``statistical'' approach to
model dwarfs.  This means that, on average, star formation in dwarfs
may proceed in the form of intermittent periods of star-forming
activity, each one lasting longer than a single \it instantaneous \rm
burst of star formation. The implication is that dwarfs should have a
sort of self-regulation mechanism capable to inhibit an effective gas
lost while keeping the low metal content. It is very plausible that
this mechanism is related to the mass of the galaxy (or the dark halo)
but also to the environment. An observational evidence of the latest
is found in the dependence of the number of early dwarf-to-giant ratio
with the richness of the environment (Ferguson \& Sandage 1991). Our
local group of galaxies is completely dwarf-dominated, with a LF
extending over more than $\sim 13$ magnitudes (van der Berg 1992). Some
of the local dwarfs are red spheroidals depleted of gas, while some
others (eg. Carina, GR8) show evidence that the star-forming activity
has been proceeding over periods of several Gyrs. Therefore, a much
more realistic model for dwarfs should account for the different
evolutionary paths that dwarfs, depending on the dark halo mass, the
environment, or both, have followed. Unfortunately little is known yet
about the star formation history in these faint, small \it (slippery)
\rm systems.

In any case the purpose of the present work is to show that the
inclusion of dwarfs is a ``key'' issue toward a proper interpretation
of the deep counts of galaxies. The success of the very simple models
in fitting both the counts in all different photometric bands and the
colors of the faint galaxies in the HDF, while giving reasonable
predictions for the redshift distributions supports this
conclusion. Moreover, it seems clear that the very faint end of the
counts is dominated by these intrinsically faint objects, i.e.
reflects the faint end of the local LF. The inclusion of dwarfs also
provides of a simple explanation to the large amount of ``weirdos''
objects observed in the deep HDF images. Nevertheless it is important
to point out that some authors (eg. Ellis et al. 1996) have suggested
that some of these peculiar systems may be sub-systems that will
eventually merge to form ellipticals, although, as already mentioned,
this merging process very likely occurred at high redshift ($z > 3$).

A further success of the models shown here comes from the fit to
the angular size distribution of the galaxies in the HDF. Though the
modeling of dwarfs is very simplistic, mainly due to the lack of
enough observational grounds, still it was shown that the predicted
angular sizes of the dwarf population fall well into the observed
range. Discrepancies between data are models were found to be due to
the ``bright'' galaxies, and in fact these are smoothed out when some
disk size evolution is introduced.  

The model of dwarfs proposed here to fit the counts, colors, redshifts
and angular size distribution of faint galaxies relies into two
parameters: the normalization (i.e. number density) of dwarfs and the
star formation history. Whereas the first one is difficult to constrain
due to the intrinsic faintness of these galaxies, the star formation
history can be tested using local dwarfs, both following the approach
by Smecker-Hane et al. of resolving individual stars but also by means
of detailed studies of the chemical abundances.

The note of ``pessimism'' is given by the fact that adding ``gentle''
dwarfs to the models makes possible to fit counts, colors and
redshifts both in the open and in the EdS case. Because the faint end
of the counts are now dominated by low-z dwarfs, for which the
normalization is, as remarked above, very uncertain (not to say totally
un-known), $q_0$ won't be easily constrain by comparing the level of
faint galaxies with the \it availability \rm of large volumes at
high-z. A way out is to measure the amount of high-z (eg. $z > 3$)
galaxies at faint levels, following for example the U-drop procedure
developed by Steidel et al.  (1995). The ratio of low-z to high-z
giants is more likely to provide a closer constrain on the value of
$q_0$.  For the I-band, where counts are comparatively deeper than in
the other bands, the open models over-predict the counts at the faint
levels while the EdS models provide a much better fit. However the
uncertainties in the modeling of dwarfs do not allow to
extract any strong conclusion from this.

Finally, it is worth noticing that testing the dwarf model,
i.e. proving by means of redshifts that the bulk of faint galaxies in
the deep counts is actually located at low-redshifts, has not only
important implications for our understanding of the galaxy formation
and evolution processes, but it provides of a further test to the
standard cosmological framework. If bright counts were dominated by
giants while the main contribution to the faint counts was given by
dwarfs at lower-redshifts, this would imply that the counts of giants
\it flatten, \rm i.e.  is leveling off as the volume elements at
high-redshifts are increasing slower than in an Euclidean ($d^3$)
geometry.

\section{Summary and Conclusions}

\begin{itemize}

\item The counts and colors of galaxies in the Hubble Deep Field can
be easily fitted by simple luminosity evolution models which
incorporate a numerous population of dwarf galaxies. Because of the
present lack of knowledge on the space density of dwarfs, reasonable
fits can equally be provided in both open and flat cosmologies. To
constrain $q_0$ the high-z to low-z bright galaxies ratio is then
needed, which requires of knowledge on the redshifts of the very faint
population.

\item The incorporation of dwarfs to the models provides of a simple
explanation to the large number of \it weirdos \rm galaxies seen in
the deep Hubble images. While E and S counts can be fitted by means of
simple pure luminosity evolution models, the high level of W counts
requires of an extra-population of dwarfs. Previous claims that the
results from the counts as a function of morphology either challenge
the morphological classification system and/or support that strong
merging is present at low-to-moderate redshifts are no longer
sustained.

\item In order to fit the counts to the very faint levels by means of
dwarfs it is neccesary to invoke that the star formation history in
dwarfs is more complex than previously thought. Models in which the
star formation takes place in single, very short episodes predict large
amounts of red remnants not seen in the HDF deep counts. By allowing
the star formation to take place over longer periods the number of
remnants is severely reduced, and the color distribution of very faint
galaxies in the HDF can be nicely fitted. Then, a kind of
self-regulation mechanism capable to keep the low metal content in
dwarfs while the star formation is lasting for longer period has to be
invoked. Observational evidence for the complexity of the evolutionary
path followed by dwarfs has been found (eg. Smecker-Hane et al.  1996)
in local dwarfs.

\item As a further test to the simple dwarf-rich models shown here the
isophotal size distribution of the HDF galaxies was compared with
predictions from the models. It was found that model-dwarf sizes are
comparable to the observed sizes of the HDF galaxies, though to fit the
distribution of the whole population it was neccesary to include some
disk size evolution for the spirals.  
\end{itemize}

\acknowledgements 

I wish to acknowledge helpful discussions with Tom Shanks, Nigel
Metcalfe, Dick Fong, and Jon Gardner. I also thank Mariano Moles for
his careful reading of this manuscript, Nigel Metcalfe for the help
with the HDF data, and the referee, Henry Ferguson, whose comments helped
to largely improved the paper.  This work was supported by an EU
fellowship, and a Spanish CSIC grant.


\begin{references}



\reference
Abraham, R.G., Tanvir, N.R., Santiago, B.S., Ellis, R.S., Glazebrook,
K. \& van der Berg, S. 1996 MNRAS, 279, L47. 


\reference
Babul, A. \& Rees, M.J. 1992, MNRAS 255, 346.

\reference
Babul, A. \& Ferguson, H. 1996, ApJ 458, 100.

\reference
Barger, A. et al. 1996. In proc. of the 37th Herstmonceux Conference ``HST and
the High Redshift Universe''. In press.

\reference
Baugh, C., Frenk, C.S.F. \& Cole, S. 1996, MNRAS, 282, 27.

\reference
Binggeli, B. 1985. In ``Star Forming Dwarf Galaxies and Related
Objects'', ed. by D. Kunth, T. Thuan and J. Van (Editions Frontieres,
Paris), p. 52. 

\reference
Binggeli, B., Sandage, A. \& Tamman, E.A. 1985, AJ 90, 1681.

\reference
Binggeli, B., Sandage, A. \& Tarenghi, E.A. 1984, AJ 89, 64.


\reference
Boroson, T.A. 1981, ApJS, 46, 177.   

\reference
Bosma, A. \& Freeman, K.C. 1993, AJ 106, 1394.

\reference
Bothun, G.D., Mould, J.R., Caldwell, N., \& MacGillivray, H.T. 1986, AJ
92, 1007. 

\reference
Bower, R.G., Lucey, J.R. \& Ellis, R.S. 1992, MNRAS 249, 589.

\reference
Broadhurst, T., Ellis, R.S. \& Glazebrook, K. 1992, Nature 355, 55. 

\reference
Bruzual, G. \& Charlot, S. 1993, ApJ 405, 538.

\reference
Caldwell, N. \& Bothun, G.D. 1987, AJ 94, 1126. 

\reference
Campos, A., Moles, M. \& Masegosa, J. 1991, AJ 106, 1784.

\reference
Campos, A. \& Shanks, T. 1996, MNRAS, in press.

\reference 
Cayon, L., Silk, J. \& Charlot, S. 1996, ApJ 467, L53.

\reference
Cole, S., Aragon-Salamanca, A., Frenk, C.F.S., Navarro, J.F. \& Zefp, S.E. 
1994, MNRAS 271, 781.


\reference
Couch, W.J. \& Newell, E.B. 1984, ApJ Suppl 56, 143.

\reference
Colless, M., Ellis, R.S., Taylor, K. \& Hook, R.N.  1990, MNRAS 244, 408.

\reference
Colless, M., Ellis, R.S., Taylor, K. \& Peterson, B.A.  1993, MNRAS 261, 19.

\reference
Cowie, L.L., Songalia, A. \& Hu, E.M. 1996 AJ, submitted.

\reference
Crampton, D., Le Fevre, O., Lilly, S.J. \& Hammer, F. 1995, ApJ 455, 108.


\reference
Davies, J.I. \& Phillipps, S. 1988 MNRAS, 233, 533.

\reference
de Jong, R. 1991, PhD Thesis, Univ. Groningen. 

\reference
Dekel, A. \& Silk, J. 1986, ApJ, 303, 39.

\reference
Dickinson, M. 1996. In proc. of the 37th Herstmonceux Conference ``HST and
the High Redshift Universe''. In press.

\reference
Djorgovski, S. et al. 1995, ApJ 438, L13. 

\reference
Driver, S.P., Phillips, S., Davies, J.I., Morgan, I. \& Disney, M.J., 1994,
MNRAS 266, 155.

\reference
Driver, S.P., Windhorst, R.A., Ostrander, E.J., Keel, W.C., Griffiths, R.E. \&
Ratnatunga, K.U. 1995, ApJ 449, 23.

\reference
Efstathiou, G., Ellis, R.S. \& Peterson, B.A. 1988, MNRAS 231, 431.

\reference
Ellis, R.S. 1983, in ``The Origin and Evolution of Galaxies''. Eds
B.J.T. Jones and J.T. Jones. Reidel:Dordrecht, p. 255.

\reference
Ellis, R.S., Smail, I., Dressler, A., Couch, W.J., Oemler, A.Jr., Butcher, H. 
\& Sharples, R.M. 1996, ApJ submitted. 


\reference 
Ferguson, H.C. \& Sandage, A. 1991, AJ 101, 765.

\reference
Franchescini, A., Mazzei, P., de Zotti, G. \& Danese, L. 1994, ApJ 427, 140.

\reference
Freeman, K.C. 1970, ApJ 160, 811.

\reference
Gardner, J.P., Sharples, R.M., Carrasco, B.E. \& Frenk, C.S.F. 1996a, MNRAS,
submitted.

\reference
Gardner, J.P.,  Shanks, T., 
Metcalfe, N., Campos, A. \& Fong, R. 1996b, in preparation.

\reference
Glazebrook, K., Ellis, R.S., Santiago, B.S. \& Griffiths, R. 1995, 
MNRAS, 275, L19.

\reference
Griffiths, R.E. et al. 1996. In proc. of the 37th Herstmonceux 
Conference ``HST and the High Redshift Universe''. In press.

\reference
Guiderdoni, B. \& Rocca-Volmerange, B. 1991, A\& A 252, 435.

\reference
Guhathakurta, P., Tyson, J.A. \& Majewski, S.R. 1990, ApJ 357, L9.

\reference
Hall, P. \& Mackay, C.D. 1984, MNRAS 210, 979. 

\reference
Holtzman et al. 1995, PASP 107, 156.

\reference
Ichikawa, S.I., Wakamatsu, K.I. \& Okamura, S. 1986, ApJS 60, 475.

\reference
Im, M., Casertano, S., Griffiths, R.E., Ratnatunga, K.U. \& Tyson,
J.A. 1995, ApJ 441, 494.


\reference
Im, M., Griffiths, R.E. \& Ratnatunga, K.U. 1997, preprint.

\reference
Impey, C., Bothun, G. \& Malin, D. 1988, ApJ 330, 634.

\reference
Infante, L., Pritchet, C. \& Quintana, H. 1986, AJ 91, 217.

\reference
Jones, L.R., Fong, R., Shanks, T., Ellis, R.S. \& Peterson, B.A. 
1991, MNRAS 249, 481.

\reference
J\o rgensen, I., Franx, M. \& K\ae rgaard, P. 1995, MNRAS 273, 1097.

\reference
Kauffmann, G., Guiderdoni, B. \& White, S.D.M. 1994, MNRAS 267, 981.

\reference
Kauffmann, G. 1996, preprint.

\reference
Kent, S. 1985, PASP 97, 165.

\reference
Koo, D.C. \& Kron, R.G. 1992, ARA\& A 30, 613.

\reference
Koo, D.C. 1986, ApJ 311, 651.

\reference
Koo, D.C. \& Kron, R.T. 1980, PASP 92, 527.

\reference
Kron, R.G. 1978, PhD thesis, Univ. California, Berkeley. 

\reference
Kron, R.G. 1987, in ``Nearly Normal Galaxies: From Planck time to 
the present''. Proc. of the Eigth Santa Cruz Summer Workshop 
in Astronomy and Astrophysics. New York: Springer-Verlag, p. 300.

\reference
Kunth, D., Maurogordato, S. \& Vigroux, L. 1988, A\& A 204, 10.

\reference
Lacey, C.G. \& Fall, S.M. 1985, ApJ 290, 154.

\reference
Lilly, S., Cowie, L.L. \& Gardner, J.P. 1991, ApJ 369, 79.

\reference
Loveday, J., Peterson, B.A., Efstathiou, G. \& Maddox, S.J. 1992,
ApJ 390, 338.

\reference
MacLeod, B.A., Bernstein, G.M., Rieke, M.J., Tollestrup, E.V., \& Fazio, G.G. 
1995, ApJ Suppl. 96, 117. 

\reference
Madau, P. 1995, ApJ 441, 18.

\reference
Maddox, S.J., Sutherland, W.J., Efstathiou, G., Loveday, J. \& Peterson, B.A. 
1990, MNRAS 247, 1p.

\reference
Masegosa, J., Moles, M. \& Campos, A. 1994, ApJ 420, 576.

\reference
Marquez, I. \& Moles, M. 1997, A\& A in press.

\reference
Marzke, R.O., Huchra, J.P., Geller, M.J. 1994, ApJ 428, 43.

\reference
Metcalfe, N., Shanks, T., Fong, R. \& Jones, L.R. 1991, MNRAS 249, 498.

\reference
Metcalfe, N., Shanks, T., Fong, R., \& Roche, N. 1995, MNRAS, 273, 257.

\reference
Metcalfe, N., Shanks, T., Campos, A., Fong, R. \& Gardner, J.P. 1996,
Nature 383, 236.

\reference
Metcalfe, N., Shanks, T. \&  Fong, R., 1997, in preparation. 

\reference
Moles, M., Aparicio, A. \& Masegosa, J. 1990, A\& A 228, 310.

\reference
Pozzetti, L., Bruzual, G. \& Zamorani, G. 1996, MNRAS in press.

\reference
Roche, N., Ratnatunga, R.E., Griffiths, R.E., Im, M. \& Neuschaefer, L.
1997, preprint. 

\reference
Sandage, A. 1961, ApJ 133, 355. 

\reference
Shanks, T. 1980. In ``The Galactic and Extragalactic Background
Radiation'', p. 269, eds. S. Bowyer and C. Leinert. Dordrecht: Kluwer.

\reference
Smail, I., Hogg, D.W., Yan, L. \& Cohen, J.G. 1996, MNRAS in press.

\reference
Smecker-Hane, T.A., Stetson, P.B., Hesser, J.E. \& van der Berg, D.A. 1996,
astro-ph/9601020. 

\reference
Soifer, B.T. et al. 1994, ApJ 420, L1.

\reference
Steidel, C.C., Pettini, M. \& Hamilton, D. 1996, AJ 110, 2519.

\reference
Tinsley, B.M. 1972, A\& A 20, 383.

\reference
Trentham, N. 1996, PhD Thesis, Univ. Honolulu.

\reference
Tyson, J.A. 1988, AJ 96, 1.

\reference
van der Berg, S. 1992, A\& A 264, 75.

\reference
Wang, B.Q. 1991, ApJ 383, L37.

\reference
Wang, B.Q. \& Silk, J. 1994, ApJ 427, 759.

\reference
Williams, R.E. et at. 1996, AJ 112, 1335. 

\reference
White, S.D.M. \& Frenk, C.S. 1991, ApJ 379, 52.




\end{references}
\end{document}